\documentstyle[aps,twocolumn,epsf]{revtex}

\begin{document}

\wideabs{
\title{Intermittent dislocation flow in viscoplastic deformation}
\author{M.-Carmen Miguel$^{1,2}$, Alessandro Vespignani$^{1}$, Stefano
Zapperi$^{3}$, J\'er\^ome Weiss$^{4}$, and Jean-Robert Grasso$^{5}$\\
$^1$The Abdus Salam International Centre for Theoretical Physics\\
P.O. Box 586, 34100 Trieste, Italy\\ $^2$ Departament de F\'{\i}sica
Fonamental, Facultat de F\'{\i}sica, Universitat de Barcelona\\
Av. Diagonal 647, 08028 Barcelona, Spain \\ $^3$INFM, Universit\`a "La
Sapienza", P.le A. Moro 2, 00185 Roma, Italy\\ $^4$LGGE-CNRS, 54 rue
Moli\'ere, B.P. 96, 38402 St Martin d'H\'eres Cedex, France\\
$^5$LGIT, B.P. 53X, 38041 Grenoble Cedex 9, France}

\maketitle

\vspace{0.5truecm}

{\bf The viscoplastic deformation (creep) of crystalline materials
under constant stress involves the motion of a large number of
interacting dislocations~\cite{Hirth92}. Analytical methods and
sophisticated `dislocation-dynamics' simulations have proved very
effective in the study of dislocation patterning, and have led to
macroscopic constitutive laws of plastic
deformation~\cite{HAN-98,ZAI-99,Kubin87,Amodeo90,Groma93,Fournet96,Gil91,Lev98}.
Yet, a statistical analysis of the dynamics of an assembly of
interacting dislocations has not hitherto been performed. Here we
report acoustic emission measurements on stressed ice single crystals,
the results of which indicate that dislocations move in a scale-free
intermittent fashion. This result is confirmed by numerical
simulations of a model of interacting dislocations that successfully
reproduces the main features of the experiment. We find that
dislocations generate a slowly evolving configuration landscape which
coexists with rapid collective rearrangements. These rearrangements
involve a comparatively small fraction of the dislocations and lead to
an intermittent behavior of the net plastic response. This basic
dynamical picture appears to be a generic feature in the deformation
of many other materials~\cite{NEU-83,Becker32,Bengus67}. Moreover, it
should provide a framework for discussing fundamental aspects of
plasticity, that goes beyond standard mean-field approaches that see
plastic deformation as a smooth laminar flow.}
}

Whenever dislocation glide is the dominant plastic deformation
mechanism in a crystalline material, we observe a constant strain-rate
regime usually described by Orowan's relation $\dot\gamma \sim \rho_m
b v$.  Here, the plastic strain-rate of the material $\dot\gamma$ is
simply related to average quantities such as $\rho_m$, the density of
mobile dislocations, and $v$, their average velocity along the slip
direction (parallel to the Burgers vector $b$)~\cite{Hirth92}.
Transmission electron micrographs of plastically deformed materials
display, on the other hand, complex features such as cellular
structures and fractal patterns~\cite{HAN-98,ZAI-99}, which are the
fingerprint of a complex multiscale dynamics not appropriately
accounted for by the mean-field character of Orowan's relation.  In
addition, rapid slip events \cite{NEU-83} have been observed in the
plastic deformation of various metals and alloys
\cite{Becker32,Bengus67}, and in the Portevin-LeChatelier
effect\cite{ANA-99}. We believe that formulating plastic deformation
as a nonequilibrium statistical mechanics problem \cite{HAN-96}
requires a substantial understanding of basic collective dislocation
dynamics.

Experimentally, the complex character of collective dislocation
dynamics can be revealed by acoustic emission measurements.  The
acoustic waves recorded in a piezoelectric transducer disclose the
pulse-like changes of the local displacements in the material during
plastic deformation, whereas a smooth plastic flow would not be
detected~\cite{ROU-83}. Thus, this method is particularly useful for
inspecting possible fluctuations in the dislocation velocities and
densities.

Ice single crystals can be used as a model material to study glide
dislocation dynamics by acoustic emission~\cite{WEI-97,WEI-00} due to
the following reasons: (i) Transparency allows direct verification
that acoustic emission activity is not related to microcracking. (ii)
Within the range of temperature and stress corresponding to our
experimental conditions, diffusional creep is not a significant
mechanism of inelastic deformation \cite{DUV-83} which, in hexagonal
ice single crystals, occurs essentially by dislocation glide on the
basal planes along a preferred slip direction. (iii) An excellent
coupling between sample and transducer can be obtained by
fusion/freezing.

\begin{figure}
\centerline{
	\epsfxsize=8.0cm 
	\epsfbox{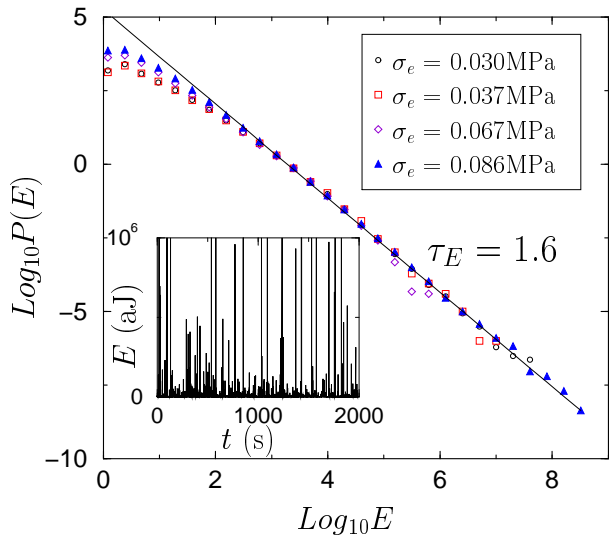}
        }
\vspace{0.25cm}
\caption{{\em Statistical properties of the acoustic energy bursts
recorded in ice single crystals under constant stress.} The main
figure shows the distribution of energy bursts for the different
loading steps. The cutoffs depend slightly on the applied stress, but
the power law exponent remains the same. The fit yields an exponent
$\tau_E=1.60\pm 0.05$. Inset, a typical recorded acoustic signal. The
creep experiment is performed at $T=-10^\circ$C under different
constant compression stresses ranging from $\sigma=0.58$MPa to
$\sigma=1.64$MPa. The stress is applied at an angle with respect to
the c-axis, giving rise to a small resolved shear stress acting across
the glide plane ($\sigma_e=0.030$MPa-$0.086$MPa). The frequency
bandwidth of the transducer was 10 to 100kHz. The amplitude range
between the minimum and the maximum recordable thresholds is 70dB,
that is, a range of seven orders of magnitude in energy. }
\label{fig1}
\end{figure}

Uniaxial compression creep experiments are performed on artificial ice
single crystals, employing several steps of constant applied
stress. We observe an intense acoustic activity, exhibiting a strong
intermittent character (see the inset of Figure~\ref{fig1}).  We
measure the energy associated to each acoustic burst and analyze its
statistical properties.  In Fig.~\ref{fig1} we show that the
probability distribution of energy burst intensities exhibits a power
law behavior spanning several decades. This fact is an indication that
a large number of dislocations move cooperatively in an intermittent
fashion. A similar behavior has been observed in the
Portevin-LeChatelier effect \cite{ANA-99}, a plastic instability where
the intermittent flow is ruled by the interaction of dislocations and
diffusing solute atoms. The intermittency observed in our case is
different as we only have interacting dislocations subject to an
external stress, without any other alien element interfering with
their dynamics.

Multiscale properties and pattern formation are ubiquitous in plastic
materials and we expect that the large dynamical fluctuations observed
in ice compression experiments are also a prevalent feature of plastic
deformation micromechanics. This general picture implies that the
experimental phenomenology should be reproducible in simple numerical
simulations of discrete dislocation dynamics that preserve the
relevant characteristics of the system under study. As in previous
dislocation dynamics models \cite{Kubin87,Amodeo90,Groma93,Fournet96},
we consider a two-dimensional cross-section of the crystal (that
is, the $xy$ plane), and randomly place $N$ straight-edge dislocations
gliding along a single slip direction parallel to their respective
Burgers' vectors ${\bf b}$ (that is, the $x$ direction). This implies
that we have point-like dislocations moving along fixed lines parallel
to the $x$ axis. This simplification effectively describes materials
like ice crystals that, owing to their strong plastic anisotropy,
deform by glide on a single slip system. An edge dislocation with
Burgers vector $b$ located at the origin gives rise to a shear stress
$\sigma_{s}$ at a point ${\bf r}=(x,y)$ of the form
\begin{equation}\label{eq:1}
\sigma_{s}({\bf r})=b\mu\frac{x(x^2-y^2)}{2\pi(1-\nu)(x^2+y^2)^2},
\end{equation} 
where $\mu$ is the shear modulus and $\nu$ is the Poisson
ratio~\cite{Hirth92}. This long-range stress is responsible for mutual
interactions among dislocations which are important in all dislocation
dynamics models~\cite{Kubin87,Amodeo90,Groma93,Fournet96}.  Under a
constant external stress $\sigma_e$, dislocation $i$ performs an
overdamped motion along the $x$ direction described by the following
equation
\begin{equation}\label{eq:2}
\eta \frac{dx_i}{dt}=
b_i(\sum_{m\neq i} \sigma_{s}({\bf r_m}- {\bf r_i}) - \sigma_{e}),
\end{equation} 
where $\eta$ is the effective friction and $b_i$ is the Burgers
vector. Throughout the simulations, we will be dealing with
dimensionless quantities after setting the distance scale $b=1$, and
the time scale $t_o\equiv \eta b/(\mu/2\pi(1-\nu))=1$.

Other essential ingredients commonly introduced in most dislocation
dynamics models~\cite{Kubin87,Amodeo90,Groma93,Fournet96} are the
mechanisms for dislocation (i) annihilation and (ii)
multiplication. (i) When the distance between two dislocations is of
the order of a few Burgers vectors, the high stress and strain
conditions close to the dislocation core invalidate the results
obtained from a linear elasticity theory (equation (\ref{eq:1})). In
these instances, phenomenological nonlinear reactions describe more
accurately the real behavior of dislocations in a crystal. In our
model, we annihilate two dislocations with opposite Burgers vectors
when the distance between them is shorter than $2b$. (ii) The
activation of Frank-Read sources~\cite{Hirth92} is accepted as the
most relevant dislocation multiplication mechanism under creep
deformation. The activation of Frank-Read sources have been observed
in ice crystals, along with more specific multiplication
processes~\cite{Petrenko94}. Because an accurate multiplication
mechanism cannot be simulated in a two-dimensional point-dislocation
model, we have implemented various phenomenological procedures. A
simple way of taking into account the new dislocation loops generated
at different sources within the crystal is the random introduction of
opposite sign dislocation pairs in the cross-section under
consideration. The rate and frequency of this creation process depend
solely on the external stress $\sigma_e$.  We have checked that other
multiplication rules in which the creation rate depends on the local
stress yield similar results.

\newpage

\onecolumn

\begin{figure}
\centerline{
        \epsfxsize=14.0cm
        \epsfbox{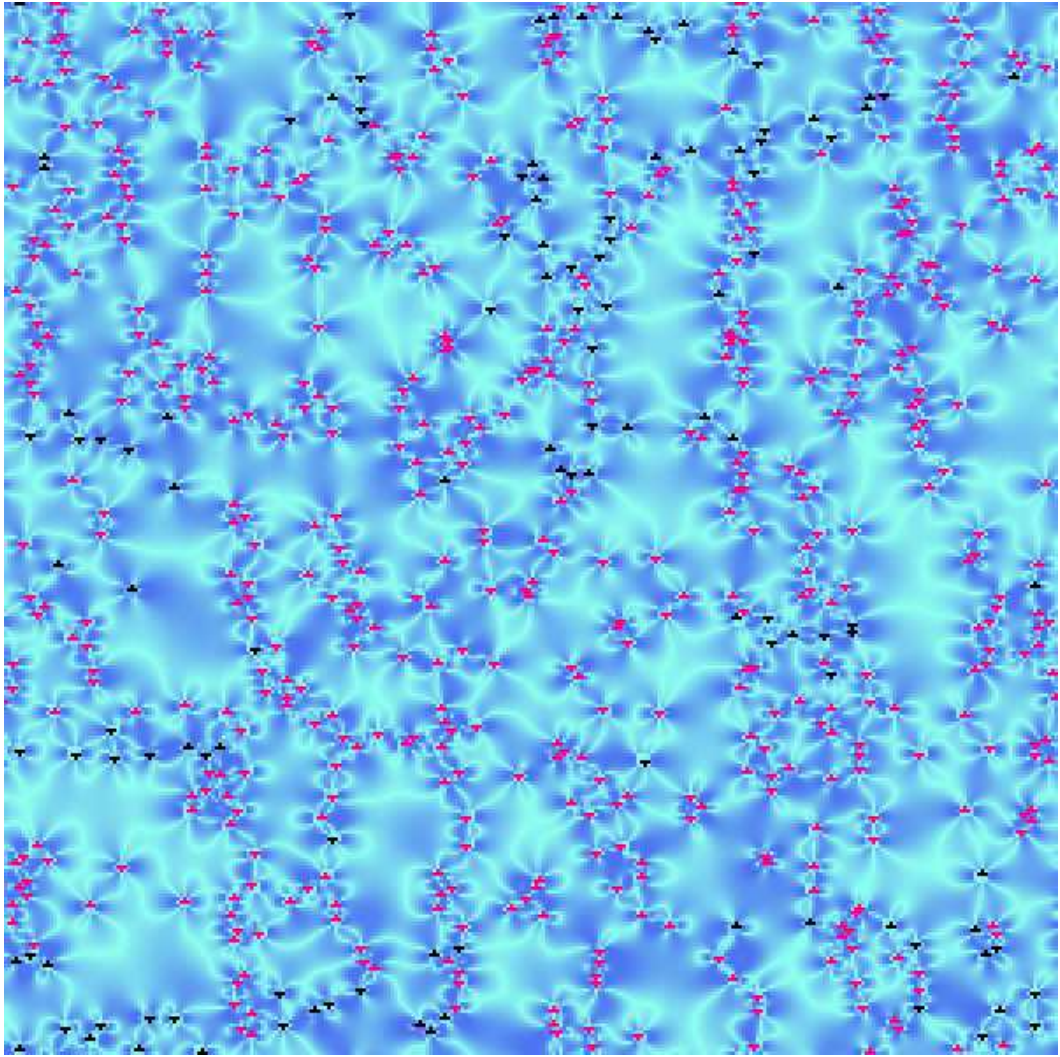}
        }
\vspace{1cm}
\caption{{\em Snapshot of the total stress field and dislocations
arrangement in a numerical simulation with $v_{\sigma}=0.025$.} We
observe metastable structure formation (dipoles and walls) and
the associated stress field which goes from light blue (low values) to
dark blue (high values). The complex low-stress pathways joining
dislocations are the result of the anisotropic elastic
interactions. Dislocations moving at low velocities (lower than
$v_{\sigma}$) are depicted in magenta, while those moving at higher
velocities are depicted in black.  Most dislocations in walls are
moving slowly, whereas isolated ones tend to move at higher speed.  In
the simulations, an initial number of $N_0=1500$ dislocations are
randomly distributed on a square cell of size $L=300$. Their Burgers'
vectors are randomly chosen to be $+b$ or $-b$ with equal
probability. In the absence of external stress, we first let the
system relax until it reaches a metastable arrangement. The number of
remaining dislocations is then $N\sim 700$. At this point, we apply a
constant shear stress and study their evolution. To avoid the
discontinuities arising from truncating long-range elastic
interactions in Eq.~\protect(\ref{eq:1}), we resort to the Ewald
summation method. We have thus exactly taken into account the
interaction of a dislocation with all the infinite periodic images of
another dislocation in the same cell.}
\label{fig2}
\end{figure}

\newpage

\twocolumn

\begin{figure}
\centerline{ 
	\epsfxsize=8.0cm 
	\epsfbox{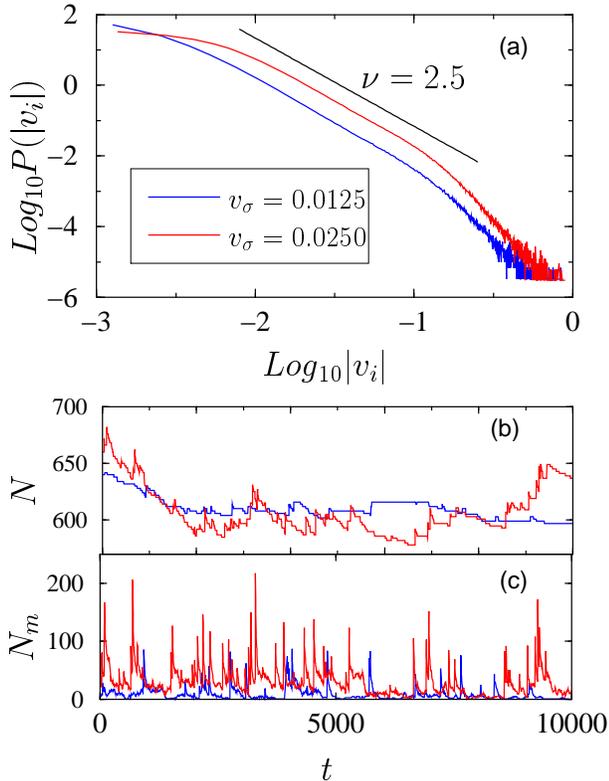} 
	}
\vspace{0.25cm}
\caption{{\em Statistical properties of dislocation velocities and
density obtained in numerical simulations.}  a) Probability density of
the individual dislocation velocities $|v_i|$ in the stationary
state. The black line is a least-squares fit of the intermediate data
points. We obtain an exponent $\nu=2.5$ for the scaling of the
intermediate velocities. b) Time evolution of the total number of
dislocations in the cell. c) Fraction of {\em fast} moving
dislocations for a given run of the simulations.}
\label{fig3a}
\end{figure}

In the initial stage of the simulations, the system relaxes slowly and
the average strain-rate decreases until it reaches a constant value
which depends on the applied stress, that is the {\em stationary creep
regime}. By monitoring the activity in a single run, we observe that
most dislocations are arranged into metastable structures (such as
walls and cells) moving at a very slow rate (see Figure~\ref{fig2} and
the Supplementary Information). A smaller fraction of dislocations,
however, move intermittently at much higher velocities, giving rise to
sudden increases of the plastic strain. In Figure~\ref{fig3a}a), we
plot the probability density of finding an individual dislocation
moving with a velocity $|v_i|$. Already at the level of individual
dislocations, the velocity distribution $P(|v_i|)$ is quite broad and
exhibits power law behavior for velocity values larger than the
external stress induced velocity $v_{\sigma}=b\sigma_e/\eta$ (we have
considered two values of $v_{\sigma}:0.0125$ and $0.025$).

\begin{figure}
\centerline{ 
	\epsfxsize=8.0cm 
	\epsfbox{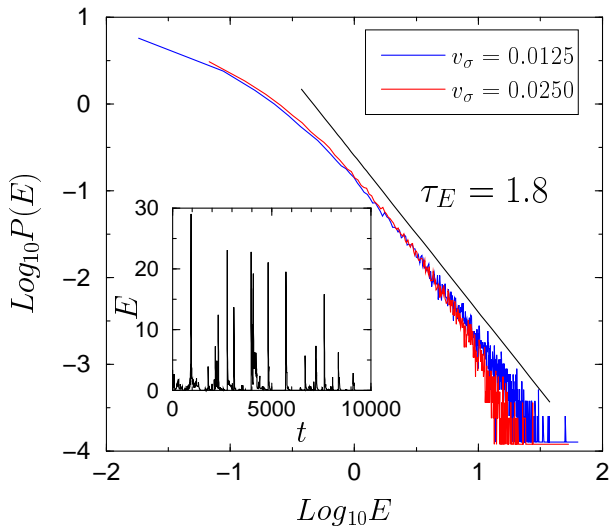} 
	}
\vspace{0.25cm}
\caption{{\em Statistical properties of the energy bursts obtained in
numerical simulations.} In the model, we define the {\em acoustic}
energy $E=V^2$ from the collective velocity $V=\sum_{i}^{'}|v_i|$ of
the {\em fast} moving dislocations.  The time evolution of $E$ for a
given simulation run is plotted in the inset.  The energy distribution
averaged over time and over a hundred realizations with different
initial conditions is plotted in the main graph. The black line
represents the best fit of the intermediate data points. We obtain an
exponent $\tau_E=1.8\pm 0.2$ for both values of the external stress.}
\label{fig3}
\end{figure}

In the presence of a large number of dislocations, the acoustic signal
that is detected experimentally is due to the superposition of the
waves emitted by each moving dislocation~\cite{ROU-83}. In particular,
it has been shown that a single dislocation performing a sudden
movement with velocity $v_0$ for a short time $\tau$ gives rise to an
acoustic wave whose amplitude is proportional to
$v_0$~\cite{ROU-83}. The high amplitude pulses detected in the
experiment cannot be ascribed to uncorrelated emissions from each
individual dislocation, but rather to the cooperative motion of
several dislocations occurring, for example, after the activation of a
Frank-Read source, or if a dislocation cluster breaks apart. To gain
further insight into this behavior, in Figures~\ref{fig3a}b) and
\ref{fig3a}c), we show the evolution of the total number of
dislocations as well as that of the {\em fast} moving dislocations,
that is, dislocations moving faster than if they were moving under the
only action of the external stress, $v_i>v_{\sigma}$ (other threshold
values yield equivalent results). After the injection of a few new
dislocations or the annihilation of a dislocation pair, several other
dislocations start to move and rearrange, not necessarily in the close
vicinity of the triggering event.  To quantify this effect, we measure
the collective velocity $V=\sum^{'} |v_i|$ of the {\em fast} moving
dislocations and define the acoustic energy as
$E=V^2$~\cite{WEI-00}. In the inset of Figure~\ref{fig3}, we see that
the signal $E(t)$ consists of a succession of intermittent and
pronounced bursts, each one signalling the onset of collective
dislocation rearrangements. The slow and continuum motion of
dislocation structures ($v_i<v_{\sigma}$) is not considered as it only
sets up a background noise signal which cannot be detected
experimentally.  In Fig.~\ref{fig3}) we show that the distribution of
energy bursts, obtained by sampling the signal over different times
and realizations, decays very slowly, spanning various decades. For
intermediate values, the distribution shows an algebraic decay with an
exponent $\tau_E=1.8\pm 0.2$, in reasonable agreement with experiments
(see Fig.~\ref{fig1}). The maximum number of dislocations we can
handle in our simulations severely restricts the maximum value of the
signal in the system. Consequently, the extension of the power law
regime grows with the number of dislocations, but is eventually
bounded by the different nature of the extremes statistics.

The broadly distributed intermittency resulting from the statistical
analysis can be interpreted as a nonequilibrium transport phenomenon
with scaling properties. Scaling behavior is usually associated with the
proximity of a critical point and is characterized by a high degree of
universality. Critical exponents only depend on a few fundamental
parameters such as the effective dimensionality, and the basic
symmetries of the system. Specifically, the reasonable quantitative
agreement between our model and the experimental data is due to the
strong plastic anisotropy present in ice single crystals, that can
thus be well described by a two dimensional model.  Although the
quantitative results we obtain should be restricted to the case of
single slip systems, the general features of the observed intermittent
flow regime appear to be generic in plastic
deformation~\cite{Becker32,Bengus67}.  The particular universality
class will depend on the effective dimensionality of each specific
material as well as the particular deformation mechanism, and one can
still expect to obtain the right critical exponents from simplified
models if the relevant symmetries are correctly taken into account.

The close-to-criticality nonequilibrium framework motivates possible
analogies with elastic manifolds driven in random media such as fluid
flow in porous media, vortices in superconductors, and charge density
waves~\cite{Kardar98}.  In all of these systems, a critical value of
the driving force separates a static regime from a moving one, and
scaling is observed only close to this point. The intermittent
behavior close to criticality is in these cases associated to static
random heterogeneities which exert space dependent pinning forces on
the moving objects.  On the contrary, the present dislocation dynamics
model, as well as the ice experiment, does not contain any quenched
source of pinning forces. Dislocation themselves, through the various
structures such as dipoles and walls, are self-generating a pinning
force landscape in which the dynamics is virtually frozen; i.e.  a
slow dynamics state. Creation and annihilation mechanisms, often
triggered by the presence of unsettled dislocations, allow the system
to jump between slow dynamics states through bursts of activity. This
behavior is reminiscent of driven-dissipative self-organizing systems
that achieve criticality in the limit of a very slow driving
\cite{Jensen98}.

The emerging scenario poses many basic theoretical questions.  Is
there a critical stress value below which the system decays into a
slow dynamics state? Can the large dynamical fluctuations be
associated with a diverging response function and thus, be related to
a critical point?  Can we derive scaling laws relating the various
observed exponents as suggested by the theory of critical phenomena?
The answers to all these questions pave the way to the nonequilibrium
statistical theory of dislocation motion that is needed for a deeper
understanding of the micromechanics of plastic deformation.

We thank R. Pastor-Satorras, M. Rub\'{\i}, A. Scala, and M. Zaiser for
useful discussions, and O. Brissaud, and F. Domin\'e for help in the
preparation of single crystals.  We acknowledge partial support from
the European Network contract on ``Fractal Structures and
Self-organization''. J. W. is supported by the "Action th\'ematique
innovante'' of INSU-CNRS. Acoustic emission monitoring devices were 
financed by Universit\'e Joseph Fourier.

\end{document}